\documentstyle[psfig]{mn}

\def\gs{\mathrel{\raise0.35ex\hbox{$\scriptstyle >$}\kern-0.6em
\lower0.40ex\hbox{{$\scriptstyle \sim$}}}}
\def\ls{\mathrel{\raise0.35ex\hbox{$\scriptstyle <$}\kern-0.6em
\lower0.40ex\hbox{{$\scriptstyle \sim$}}}}
\def\ls{\mathrel{\hbox{\rlap{\hbox{\lower4pt\hbox{$\sim$}}}\hbox{$<$}}}}
\def\gs{\mathrel{\hbox{\rlap{\hbox{\lower4pt\hbox{$\sim$}}}\hbox{$>$}}}}

\def\mnras {{\sc MNRAS}}

\def\nat {Nature}

\def\mpc {${\rm h^{-1}}$ Mpc}

\title[Abundance and Distribution of Filaments in 2dFGRS]
      {Inter-cluster Filaments of Galaxies Programme: Abundance
and Distribution of Filaments in the 2dFGRS Catalogue}
\author[K.\,A.\ Pimbblet, M.\ J.\ Drinkwater \& M.\ C.\ Hawkrigg]
       {Kevin A.\ Pimbblet$^1$, Michael J.\ Drinkwater and Mary C.\ Hawkrigg
        \vspace*{1mm}\\
        Department of Physics, University of Queensland, Brisbane,
        4072 Queensland, Australia\\
        $^1$ pimbblet@physics.uq.edu.au}

\date{Accepted ... ; Received ... ; in original ...}

\pagerange{000--000}

\begin{document}

\maketitle

\begin{abstract}
Filaments of galaxies are known to stretch between galaxy 
clusters at all redshifts in a complex manner.  
In this Letter, we present an analysis of the frequency 
and distribution of inter-cluster galaxy filaments selected
from the 2dF Galaxy Redshift Survey.  
Out of 805 cluster-cluster pairs, we find 
at least 40 per cent have bone-fide
filaments.  We introduce a filament
classification scheme and cast the filaments into several types 
according to their visual morphology:
straight (lying on the cluster-cluster axis; 37 per cent),
warped or curved (lying off the cluster-cluster axis; 33 per cent),
sheets (planar configurations of galaxies; 3 per cent),
uniform (1 per cent)
and irregular (26 per cent)
We find that straight filaments are more likely to
reside between close cluster pairs and they become more
curved with increasing cluster separation.  This curving
is toward a larger mass concentration in general.  
We also show that the more massive a cluster is, the more 
likely it is to have a larger number of filaments.
Our results are found to be consistent with 
a $\Lambda$ cold dark matter cosmology. 
\end{abstract}


\section{Introduction}

The filamentary structure of the Universe has long been 
predicted by structure formation modelling (e.g.\ Zeldovich, Einasto
\& Shandarin 1982; Katz et al.\ 1996; Jenkins et al.\ 1998).
In such N-body simulations, clusters of galaxies
reside at the nodes of the network of matter.
Filaments of galaxies (FOGs) themselves
are observed to stretch between clusters,
and indeed superclusters of galaxies, at low redshifts
(e.g.\ Kaldare et al.\ 2003; Einasto et al.\ 1997; 
Kalinkov \& Kuneva 1995), 
forming a characteristic sponge-like structure through the 
Universe (Erdo{\u g}du et 
al.\ 2004; Drinkwater 2000; Bond, Kofman \& Pogosyan 1996).
Moreover, it is clear from the work of Colberg et al.\ (1999) that
FOGs are very important for the baryonic mass budget of the Universe
as they can contain up to 40 per cent of the total cluster mass at 
clustocentric radii of 4--6.5 \mpc \ (we use $H_0 = 100 \ 
h$ km s$^{-1}$\,Mpc$^{-1}$ and $q_o=0.5$ throughout this work). 

Various observational campaigns are underway that
are reinforcing this web-like view of the 
Universe (Colless et al.\ 2001; 
York et al.\ 2000; Einasto et al.\ 2001).
By examining the regions between close cluster pairs,
observational evidence for many
and varied FOGs is growing 
(e.g.\ Pimbblet \& Drinkwater 2004; Dietrich et al.\ 2004;
Gal \& Lubin 2004; Ebeling, Barrett, \& 
Donovan 2004; Drinkwater et al.\ 2004, amongst others)
and not only at optical wavelengths (e.g.\ 
Durret et al.\ 2003; Bagchi et al.\ 2002;
Tittley \& Henriksen 2001;
Ensslin et al.\ 2001; Scharf et al.\ 2000);
although detection of (X-ray emitting gas from) 
filaments has not been without
some failures (Briel \& Henry 1995).
One is left considering several questions:
(i) how common are FOGs?; 
(ii) how surprised should one be to
find a filament of a given length or morphology?

In a study designed to address these questions
Colberg, Krughoff \& Connolly (CKC; 2004) investigate the
frequency and distribution of filaments in a $\Lambda$ cold
dark matter ($\Lambda$CDM)
Universe using simulations from Kauffmann et al.\ (1999).
They show that approximately half of all inter-cluster filaments
are warped (lying off the cluster-cluster axis) and are
statistically longer than straight filaments.  Further, FOGs are
more likely to be found between clusters that are 
spatially close and more massive clusters possess more 
filaments.

Motivated by CKC,
in this Letter, we utilize the 2dF
Galaxy Redshift Survey (2dFGRS; e.g.\ Colless et al.\ 2001) 
final data release (FDR)
to characterize a spectroscopic sample of inter-cluster filaments
and compare this distribution to CKC.  
The format of this paper is as follows.  In Section~2 we define the
filament sample from the 2dFGRS FDR.  We then 
visually type our filaments into
a new classification scheme that we introduce.
In Section~3 we investigate the fractional abundance
of different types of filament, the likelihood of finding
different types of filament connecting cluster pairs and
the average number of filaments per cluster.
Our results and caveats are then summarized in Section~4.

\section{Methodology}
The observations made by 2dFGRS are summarized by Colless et
al.\ (2001) and here we only recount the pertinent detail.
The input catalogue for 2dFGRS is
the APM survey of Maddox et al.\ (1990a,b).
Targets are selected to be
brighter than an extinction-corrected magnitude 
limit of $b_J=19.45$ within three strips of the APM survey
(NGP, SGP and random fields) covering an area in excess
of $1500$ deg$^2$.
Subsequently, quality (quality$\geq3$; see Colless et al.\ 2001) 
redshifts for 221414 galaxies have been published as part of 
the 2dFGRS FDR.
The redshift completeness of the 2dFGRS FDR is estimated to be
90 per cent ($b_J<19.0$)
with an rms redshift error of $\Delta cz = 85$ kms$^{-1}$
and a median redshift of $<$$z$$>$$ = 0.1$ (Colless et al.\ 2001).

From an earlier sample of 173084 galaxies, De Propris et al.\ (2002)
generate a catalogue of galaxy clusters and cross-correlate
them with those of Abell (Abell 1958; Abell, Corwin \& Olowin 1989),
the Edinburgh-Durham Cluster Catalogue (EDCC; Lumsden et al.\ 1992) 
and the APM survey itself (APMCC; Dalton et al.\ 1997).
De Propris et al.\ (2002) 
report over 800 individual cluster correlations 
and calculate new velocity dispersions for them.
From their catalogue, we select potential inter-cluster filaments 
to study by applying the following criteria:

\begin{itemize}
\item The clusters must be spatially close, $<10.0$ degrees on the sky.
At a median redshift of $<$$z$$>$$ = 0.1$, this corresponds to a projected
length of $\approx 45$ \mpc.

\item Their recession velocities must not differ by 
more than $\Delta cz = 1000$ kms$^{-1}$.

\item Clusters common to the Abell, EDCC and APMCC catalogues
are removed to prevent self-pairing.
\end{itemize}

Applying these criteria spawns 805 unique potential 
inter-cluster filaments.

\begin{figure*}
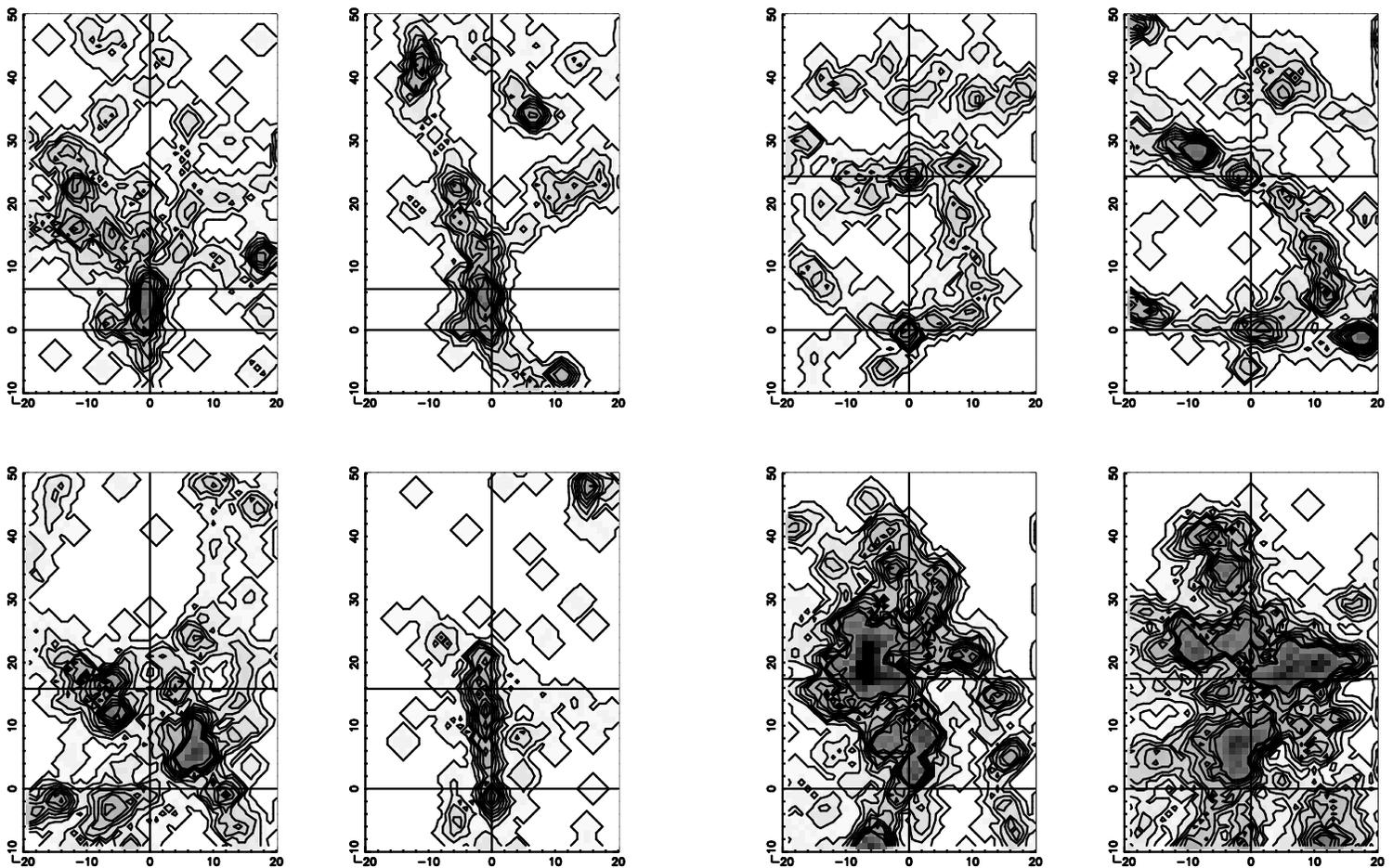

\centerline{
\psfig{file=type1.ps,angle=0,width=3.5in}
\hspace*{0.7in}
\psfig{file=type2.ps,angle=0,width=3.5in}
}
\vspace*{0.3in}
\centerline{
\psfig{file=type3.ps,angle=0,width=3.5in}
\hspace*{0.7in}
\psfig{file=type4.ps,angle=0,width=3.5in}
}
  \caption{\small{Orthogonal pairs of projected galaxy density
for selected examples of
Type I with overlapping cluster infall zones
(top left), Type II (top right), Type III (bottom left)
and Type IV--V (bottom right) FOGs. The outermost contour denotes 1 galaxy
per \mpc \ and each contour inward is an increase of 2 galaxies per \mpc \ 
to a maximum of 19 galaxies per \mpc.
The vertical solid line running up the centre of these plots is the 
inter-cluster axis.  The intersection of the horizontal solid lines 
and the inter-cluster axis denote the 
(sometimes ill-defined) locations of the cluster centres. 
All units are in \mpc \ and the depth of each plane is 10 \mpc.
}}
  \label{fig:type}
\end{figure*}

Based upon CKC, we now proceed to classify these potential 
FOGs into various types.  
Firstly, we convert all measurements into \mpc.
We then draw a vector from one cluster centre to the other
and extract all galaxies within 5 \mpc \ of this axis.  
These galaxies are then placed onto two orthogonal planes 
containing the inter-cluster axis and
smoothed with a circular top-hat function of radius 1 \mpc.
We (KAP \& MJD) then visually
inspect the two orthogonal projections of the galaxy distribution
and classify any filament(s) according to the scheme presented in
Table~\ref{tab:type}.  
Our scheme attempts to tidy up the definitions used by CKC whilst
the number of categories used is determined subjectively by
the number of distinct FOG types suggested by the data. 
After a first pass, we determined that we were missing  
highly curved filaments (specifically ones that
stretch out beyond 5 \mpc \ from the inter-cluster axis)
particularly at large cluster-cluster separations.
This prompted us to extend our cut-off width when looking at the 
orthogonal planes to $\pm$ 20 \mpc \ from the
inter-cluster axis (but retaining a depth of $\pm$5 \mpc : 
doing so yielded more Types II--V FOGs, 
especially at larger inter-cluster separations.
Typical examples for filaments of Types I through IV--V are
displayed in Figure~\ref{fig:type}.
Although we tried to automate the typing process, there are
too many (and complex) deviations from the 
simple configurations described in Table~\ref{tab:type} to reliably
employ any automated process (c.f.\ CKC who experience the same problem).
Ideally, we would replace our subjectivity with new, objective 
statistics of FOG structure (Bharadwaj \& Pandey 2004), 
but this is beyond the scope of the present work.

%
%
\begin{table*}
\begin{center}
\caption{The classification scheme used in this work.
\hfil}
\begin{tabular}{ll}
\noalign{\medskip}
\hline
Type & Filament Description \\
\hline
0   & {\bf Near-coincident clusters}.  The cluster pair overlaps to
such a degree that any filament present cannot be isolated.\\
I   & {\bf Straight}.  The filament of galaxies runs along the axis
from one cluster centre to the other. \\
    & At small separations, the infall regions of the clusters likely
overlap.\\
II  & {\bf Warped (Curved)}.  The galaxies lie off the axis and 
continuously curve (in a `C' or `S'-shape for example) from \\
    & one cluster centre to the other.  \\
III & {\bf Sheet (Planar; Wall)}.  The filament appears as Type I or II viewed
from one direction but are the galaxies are \\
    & approximately evenly spread out in the orthogonal view.\\
IV  & {\bf Uniform}.  Galaxies fill the space between the clusters
in an approximately uniform manner viewed from any direction.\\
V   & {\bf Irregular (Complex)}.  There are one or more connections between 
both cluster centres, but the connections\\
 & are irregular in shape and often have large density fluctuations.\\
\hline
\noalign{\smallskip}
\end{tabular}
  \label{tab:type}
\end{center}
\end{table*}

%
%

\section{Results and Discussion}

The fractions of each type of filament are presented
in Table~\ref{tab:frac}.
At least $>30$ per cent of the cluster pairs in our sample have
no filamentary connection between them.  
This percentage is much smaller than the 81 per cent 
noted by CKC.  This is likely due to:
(i) differences in what we classify as a connected cluster pair;
(ii) selection effects:
CKC only use Abell richness class $R=0$ clusters (Abell 1958)
in their analysis whereas our 2dFGRS sample contains much 
richer clusters and may therefore be more likely to display
filamentary connections;
(iii) they intentionally avoid clusters pairs that have 
tertiary clusters near the cluster axis.
(iv) we probe a larger width from the inter-cluster
axis (see Section~2).

%
%
\begin{table*}
\begin{center}
\caption{Filament percentages for different samples.
The column headed `nil' indicates that no filament
is detected.  Where present, the number in brackets is the percentage of uncertain
classifications within each type.
The `Connected' sample removes the `nil' detections from the whole sample.
Similarly; `Filaments' removes Type 0 from the `Connected' sample and
`Certain Filaments' removes all the uncertain classifications from the `Filaments' sample.
\hfil}
\begin{tabular}{lccccccc}
\noalign{\medskip}
\hline
Sample & \multicolumn{7}{c}{Percentage by Type} \\
       & 0 & I & II & III & IV & V & nil \\
\hline
Whole Sample & 6.2 (0) & 20.8 (16.5) & 22.0 (32.7) & 3.9 (56.0) & 1.9 (75.0) & 14.7 (22.7) & 30.2 (0) \\
Connected    & 8.9 (0) & 28.4 (16.5) & 32.1 (32.7) & 5.8 (56.0) & 2.7 (75.0) & 21.5 (22.7) & n/a \\
Filaments    & n/a     & 31.3 (16.5) & 35.2 (32.7) & 6.3 (56.0) & 2.9 (75.0) & 23.7 (22.7) & n/a \\
Certain Filaments & n/a & 36.9       & 33.5        & 3.6        & 0.8        & 25.9 & n/a \\
\hline
\noalign{\smallskip}
\end{tabular}
  \label{tab:frac}
\vspace*{-0.3in}
\end{center}
\end{table*}

To compare our fractions with CKC, we firstly
note that their classifications do not quite have a one-to-one
correspondence with ours.  CKC type filaments as:
straight, off-centre, warped/irregular and other.
Whilst their definition of a straight filament is identical
to our Type I, we have re-arranged their `off-centre' and
`warped/irregular' into Types II and V.  Their `other' 
configurations, however, include definitions of our Types III 
and IV, so a direct comparison (albeit as a fraction of the 
whole sample) is possible.
In Table~\ref{tab:compare}, we compare the overall fractions
of our sample (Table~\ref{tab:frac}) to the CKC fractions.
There is an excellent similarity between the percentages
for all filament types.  
We note that accounting for probing a larger width
from the inter-cluster axis than CKC does not 
significantly change the relative overall
fractions in Table~\ref{tab:frac}.
We conclude that the distribution of filament morphology 
is the same as found by CKC.  This suggests that the 
cluster richness differences in our two samples are
not driving the form of the filamentary connections:
Type III and IV filaments are equally rare between 
all richness classes.

\subsection{Filament length}

\begin{figure}
\centerline{\psfig{file=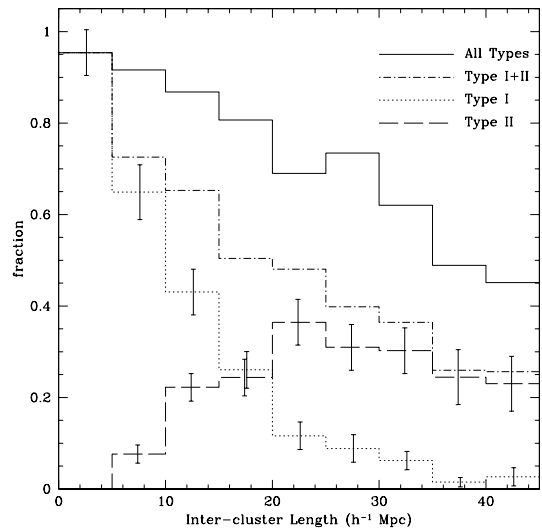,angle=0,width=3.in}}
  \caption{Abundance of filaments as a function of
cluster separation (solid line).
The dotted and dashed lines show
show the individual contributions 
of Types I and II (respectively) to the abundance
whilst the dot-dash line is the combined abundance
of Types I plus II.
Poissonian errorbars are given for Types I and II.
}
\label{fig:abund}
\end{figure}

For Type I and II filaments, we are able to
measure their length ($\approx$ inter-cluster
separation) 
and calculate their fractional abundance in the whole sample
as a function of cluster separation.  
Figure~\ref{fig:abund} displays the result of this abundance
analysis.

Very close cluster pairs, within 5 \mpc \ of each other,
will always possess a filament.  Such a filament will
nearly always be a Type I (if it is discernable from a Type 0).
This is of little surprise
given that such close cluster pairs will typically possess
overlapping infall regions (CKC; Rines et al.\ 2003; 
Diaferio \& Geller 1997).  With increasing separation,
the likelihood of being connected by a Type I or II filament 
drops $\sim$ linearly.  
However, straight filaments of Type I are much more likely
to be extant in close cluster pairs than Type II 
(Figure~\ref{fig:abund}).
As with CKC, our visual inspections of the longer Type II
filaments indicate that they are often tidally arched toward
secondary masses.
%
%
\begin{table}
\begin{center}
\caption{Comparison of our filament fractions (rounded) to 
CKC.  Quoted errors are simple Poissonian ones.  
The percentages are relative to the `certain filaments'
or `whole sample' (Table~\ref{tab:frac}) as quoted by CKC. \hfil}
\begin{tabular}{lccc}
\noalign{\medskip}
\hline
Type & Sample & CKC (\%) & This work (\%) \\
\hline
I    & certain & 38$\pm$4 & 37$\pm$3 \\
II+V & & 62$\pm$5 & 63$\pm$3 \\
\hline
III  & whole & 2$\pm$1  & 3$\pm$1 \\
IV   & & 3$\pm$1  & 2$\pm$1 \\
\hline
\noalign{\smallskip}
\end{tabular}
  \label{tab:compare}
\end{center}
\vspace*{-0.1in}
\end{table}

We note that there are differences in comparison to CKC.
At just over 40 \mpc \ inter-cluster separation, we have
a Type II fraction of $\approx0.23$ (Figure~\ref{fig:abund})
compared to $\approx0.09$ by CKC.  Since we have a larger
search radius from the inter-cluster axis than CKC, we find
more Type II filaments at these radii (see Section~2) than CKC.  
Restricting ourselves to smaller radii from the inter-cluster
axis reduces the number of long Type II filaments and
we are able to recover (within error) the distribution 
presented by CKC.

\subsection{Cluster connections}

How inter-connected are the clusters 
and how many filaments can we expect a given cluster to have?  
We can address this question by computing the average number
of filaments per cluster as a function of cluster velocity
dispersion. 
However, due to the geometry of 2dFGRS, clusters located at
the edges of the observed strips may have a smaller number
of filaments than those in the centre of the strips.  To
attempt to alleviate this, we only consider those clusters 
inside a 10 \mpc \ buffer from the observational edges of 2dFGRS.
Further, we cull from the sample high redshift
clusters ($cz>45000$ ${\rm kms^{-1}}$) as the cluster sample
is more likely to be incomplete at such redshifts.
Finally, if a given cluster is connected to two or
more clusters \emph{by the same filament}, we only count 
that filament once.

\begin{figure}
\vspace*{-1.0in}
\centerline{\psfig{file=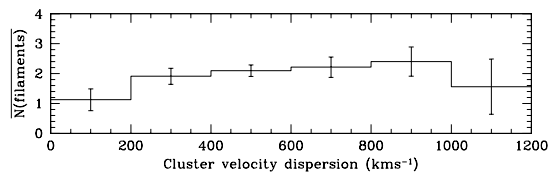,angle=0,width=3.in}}
\vspace*{-1.0in}
  \caption{Histogram of the average number of filaments
per cluster as a function of cluster
velocity dispersion with $1\sigma$ error
bars.  There is a trend for the number of filaments 
to increase with velocity dispersion.
}
\label{fig:fpv}
\end{figure}

The result of this analysis is shown in Figure~\ref{fig:fpv}.
There is a clear trend for clusters with
larger velocity dispersions to possess more filaments.
Since the virial mass of a cluster, 
$M_{virial} \propto$ (velocity dispersion)$^{2}$
(e.g.\ Binney \& Tremaine 1987), we can inductively 
state that more massive clusters are more likely to 
have more filaments.
This is consistent with $\Lambda$CDM cosmology
as more massive objects are more clustered than lower
mass objects (c.f.\ CKC who obtain a similar result).

%
%
%

%
%
%

\section{Caveats and Summary}
Clearly, our classifications of the filaments are subjective. 
For example, in $\sim 15$ per cent of cases 
we are confident that a filament conjoins the cluster pair, but
are unable to agree upon (or discern)
a definite typing according to Table~\ref{tab:type}.  
Further, the inter-cluster filament can be a small 
segment or chord of a much larger FOG that connects more than two galaxy 
clusters.  As such, it is possible that the inter-cluster FOG
may locally be straight (Type I), but on larger scales
distinctly curved (Type II).  
Next, the APM catalogue is known to be 
incomplete (Cross et al.\ 2004; Pimbblet 
et al.\ 2001), but since the missing galaxies are not preferentially
situated near clusters (Pimbblet et al.\ 2001), this should not affect
our results significantly save for lowering the galaxy density.   
Given the preceding, we also assume that under-sampled 
regions within the main strips of 2dFGRS (Norberg et al.\ 2002) 
should have the similar (small) effect of reducing the overall galaxy
density but do not present and great
threat to identifying and classifying FOGs.

We attempt to assess the impact of these factors by repeating
our experiment but restricting ourselves to galaxies brighter
than $b_J = 19.3$; $\approx3\sigma$ away from the APM cut-off 
magnitude limit (Pimbblet et al.\ 2001; Colless et al.\ 2001).  
The result of this is to increase the number of uncertain classifications
and the number of `nil' entries by $\sim$ few per cent
compared to the values presented in Table~\ref{tab:frac}. 
The relative fractions of `certain
filaments' (Table~\ref{tab:frac}), remain statistically the same, however.
We therefore do not view that this assumption has a great effect on the
results presented here.

In our selection criteria, we extract cluster pairs with 
$\Delta cz < 1000$ kms$^{-1}$.  In doing so, we may have eliminated
some pairings (i.e.\ potential filaments within a small solid angle 
along the line of sight) and effects caused by `finger-of-god' 
elongations in redshift space (see Hawkins et al.\ 2003 and
references therein).
An examination of the distribution of angles
to the line of sight, however, shows no evidence that we have 
have misclassified `fingers-of-god' as FOGs.

Our results enforce the view that cluster of galaxies are not
simple, isolated objects but nodes along a filament of galaxies
that can be made up of many sub-entities (e.g.\ other clusters).
Our main results are:

\begin{itemize}
\item We introduce a new classification scheme for
FOGs that stretch between galaxy clusters.
This scheme is based upon the different types of filament
observed in 2dFGRS and we emphasize that it is a 
purely \emph{visual} classification scheme.

\item Filamentary connections between galaxy clusters are
common at the median 2dFGRS redshift of $<$$z$$>$$=0.1$.  
In close cluster pairs, there is almost always
a Type I filament (Straight) connecting them.  At larger
cluster separations ($>5$ \mpc), Type II filaments (Curved)
become more common.  Type II filaments are often seen
to bend toward other mass concentrations nearby.

\item Whilst Type I, II and V (Irregular/Complex) filaments
are very common, Types III (Sheets) and IV (Uniform) are
very rare, comprising no more than 4 per cent of the total
filament population.  

\item The vast majority of all clusters have a filamentary
connection with their (close) neighbours.  The number of
filaments per cluster scales with the velocity dispersion, and
hence mass, of a given cluster.
\end{itemize}

All of these results are consistent with a $\Lambda$CDM cosmology
(CKC).
In the future, it will be interesting to compare these results to 
the Sloan Digital Sky Survey (e.g. York et al.\ 2000) and
on-going higher redshift surveys such as the Luminous Red Galaxy Survey
(e.g. Padmanabhan et al.\ 2004; Cannon et al.\ 2003).  
We can then address the interesting 
question of if FOGs have evolved significantly over a large
cosmologically significant timescale (e.g.\ are Type II filaments
more abundant at lower redshifts?).
We also plan to investigate alternative statistical techniques
that may provide more objective, quantitative 
measurements of filamentary structure in datasets such as this.

This work follows Pimbblet \& Drinkwater (2004) 
and is the second publication in a series
on inter-cluster filaments of galaxies.
\vspace*{-0.2in}

\section*{Acknowledgments}
We thank the 2dFGRS team for their dedication in compiling the 
the data used in this analysis; inparticular, we warmly thank 
Warrick Couch for numerous useful conversations.
We also thank the referee, Volker M\"{u}ller,
for his prompt and insightful comments that have
improved this work.
KAP acknowledges support from an EPSA University of Queensland Research
Fellowship and (in part) a UQRSF grant.

\end{document}